\documentclass[a4paper,12pt]{article}
\usepackage{epsfig}
\usepackage{amssymb}
\usepackage{amsfonts}
\usepackage{amsmath}
\usepackage{mathtools}
\usepackage{euscript}
\usepackage{verbatim}
\usepackage{latexsym}
\usepackage{graphicx}

	\usepackage{pgf,tikz}
	\usepackage{mathrsfs}
	\usetikzlibrary{arrows}
	
	\usepackage{subfigure}
	\usepackage{wrapfig}
	\usepackage[T1]{fontenc}
	\usepackage{mathtools}
	\usepackage{float}

\usepackage{tikz}
\usetikzlibrary{shapes.geometric, arrows,patterns,snakes}
\tikzstyle{ellip} = [ellipse, minimum width=3cm, minimum height=1cm,text centered, draw=black]

\newskip\humongous \humongous=0pt plus 1000pt minus 1000pt

\newif\ifdtup

\catcode`\@=11

\@addtoreset{equation}{section}

\def\@normalsize{\@setsize\normalsize{15pt}\xiipt\@xiipt
\abovedisplayskip 14pt plus3pt minus3pt%
\belowdisplayskip \abovedisplayskip
\abovedisplayshortskip \z@ plus3pt%
\belowdisplayshortskip 7pt plus3.5pt minus0pt}

\def\small{\@setsize\small{13.6pt}\xipt\@xipt
\abovedisplayskip 13pt plus3pt minus3pt%
\belowdisplayskip \abovedisplayskip
\abovedisplayshortskip \z@ plus3pt%
\belowdisplayshortskip 7pt plus3.5pt minus0pt
\def\@listi{\parsep 4.5pt plus 2pt minus 1pt
     \itemsep \parsep
     \topsep 9pt plus 3pt minus 3pt}}

\relax

\catcode`@=12

\catcode`\@=11

\def\section{\@startsection{section}{1}{\z@}{3.5ex plus 1ex minus
   .2ex}{2.3ex plus .2ex}{\large\bf}}

\def\SymBoxes#1#2#3#4{\newdimen\un@t \un@t#3%
\raisebox{#1}{\rule{#2\un@t}{#4}\hskip-#2\un@t
\@tempdimb\un@t \advance\@tempdimb by-#4\@tempcntb#2\relax%
\@whilenum{\@tempcntb>0}\do{
\rule{#4}{\un@t}\hskip\@tempdimb \advance\@tempcntb by\m@ne}%
\hskip-#2\un@t \rule[\un@t]{#2\un@t}{#4}%
\rule[\un@t]{#4}{#4}\hskip-#4
\rule{#4}{\un@t}}\hskip-#4}                

\begin{document}

\newcommand{\beq}{\begin{equation}}
\newcommand{\eeq}{\end{equation}}
\newcommand{\bea}{\begin{eqnarray}}
\newcommand{\eea}{\end{eqnarray}}
\newcommand{\beas}{\begin{eqnarray*}}
\newcommand{\eeas}{\end{eqnarray*}}
\newcommand{\defi}{\stackrel{\rm def}{=}}
\newcommand{\non}{\nonumber}
\newcommand{\bquo}{\begin{quote}}
\newcommand{\enqu}{\end{quote}}
\renewcommand{\(}{\begin{equation}}
\renewcommand{\)}{\end{equation}}
\def \eqn#1#2{\begin{equation}#2\label{#1}\end{equation}}

\def\pp{\partial_{\phi}}
\def\e{\varepsilon}
\def\Tr{ \hbox{\rm Tr}}
\def\H{ \hbox{\rm H}}
\def\HE{ \hbox{$\rm H^{even}$}}
\def\HO{ \hbox{$\rm H^{odd}$}}
\def\K{ \hbox{\rm K}}
\def\Im{ \hbox{\rm Im}}
\def\Ker{ \hbox{\rm Ker}}
\def\const{\hbox {\rm const.}}
\def\o{\over}
\def\im{\hbox{\rm Im}}
\def\re{\hbox{\rm Re}}
\def\bra{\langle}
\def\ket{\rangle}
\def\Arg{\hbox {\rm Arg}}
\def\Re{\hbox {\rm Re}}
\def\Im{\hbox {\rm Im}}
\def\exo{\hbox {\rm exp}}
\def\diag{\hbox{\rm diag}}
\def\longvert{{\rule[-2mm]{0.1mm}{7mm}}\,}
\def\a{\alpha}
\def\dag{{}^{\dagger}}
\def\tq{{\widetilde q}}
\def\p{{}^{\prime}}

\def\hsp{,\hspace{.7cm}}

\def\br{\nonumber\\}

\def \eqn#1#2{\begin{equation}#2\label{#1}\end{equation}}

\definecolor{qqqqff}{rgb}{0.,0.,1.}

\newcommand{\M}{\ensuremath{\mathcal{M}}
                    }
\newcommand{\oc}{\ensuremath{\overline{c}}}
\begin{titlepage}
\begin{flushright}
CHEP XXXXX
\end{flushright}
\bigskip
\def\thefootnote{\fnsymbol{footnote}}

\begin{center}
{\Large
{\bf Contractions from Grading \\ 
\vspace{0.1in} 
}
}
\end{center}

\bigskip
\begin{center}
{
Chethan KRISHNAN$^a$\footnote{\texttt{chethan.krishnan@gmail.com}}, Avinash RAJU$^a$\footnote{\texttt{avinashraju777@gmail.com}}}
\vspace{0.1in}

\end{center}

\renewcommand{\thefootnote}{\arabic{footnote}}

\begin{center}

$^a$ {Center for High Energy Physics,\\
Indian Institute of Science, Bangalore 560012, India}\\

\end{center}

\noindent
\begin{center} {\bf Abstract} \end{center}
We note that large classes of contractions of algebras that arise in physics can be understood purely algebraically, via identifying appropriate $\mathbb{Z}_m$-gradings (and their generalizations) on the parent algebra. This includes various types of flat space/Carroll limits of finite and infinite dimensional (A)dS algebras, as well as Galilean and Galilean Conformal algebras. Our observations can be regarded as providing a natural context for the Grassmann approach of arXiv:1312.2941. We also introduce a related notion, which we call partial grading, that arises naturally in this context.

\vspace{1.6 cm}
\vfill

\end{titlepage}

\setcounter{footnote}{0}

\section{Introdution}


It was noticed by Inonu and Wigner long ago that certain irreversible scaling limits of Lie algebras can lead to new Lie algebras. This process is called \textit{contraction} (or \textit{Inonu-Wigner contraction}) and a classic example of this is the contraction of Poincare algebra to the Galilean algebra \cite{Inonu:1953sp}. Contractions can relate symmetry algebras of limiting theories to the symmetry algebras of the original theory. This has had some applications in the context of holography.

In the context of $2+1$-dimensional gravity, Inonu-Wigner contractions have been fruitful in exploring flat space gravity and higher spin theories as the limiting cases of situations with non-vanishing cosmological constant \cite{Barnich:2006av,Barnich:2012aw,Gonzalez:2013oaa,Costa:2013vza,Fareghbal:2013ifa,Krishnan:2013wta,Basu}. It was shown that the (supersymmetric) asymptotic symmetry algebras of AdS$_3$ and higher spin theories with Brown-Henneaux-like \cite{Brown:1986nw} boundary conditions go over to (super) $\mathfrak{bms}_3$ and its higher spin generalizations \cite{Barnich:2006av,Gonzalez:2013oaa,Afshar:2013vka,Caroca:2017onr,Banerjee:2017gzj,Oscar}. Variations of this theme in a generalized AdS$_3$ context were undertaken in \cite{Stefan,Grumiller:2017sjh}. In a different context, Galilean versions of conformal algebras were obtained as non-relativistic limits of the conformal algebra in \cite{Bagchi:2009my,Martelli:2009uc}.

In \cite{Krishnan:2013wta}, a purely algebraic realization of Inonu-Wigner contraction for AdS$_3$ gravity and higher spin theories was introduced using a Grassmann approach. This prescription, apart from contracting the algebra, also worked at the level of solutions, charges, asymptotic symmetry algebras and holonomies. The prescription consists of simply replacing the inverse AdS radius $1/l$ with a Grassmann parameter $\epsilon$ and setting $\epsilon^2 =0$. 

In this work, we will show that this algebraic approach can be generalized in a few different ways by identifying and exploiting $\mathbb{Z}_n$-{\em grading} structures (and their generalizations) on the parent algebra and introducing (higher) nilpotent parameters. In fact, as we will show, many of the distinct contractions that have shown up in the physics literature over the years can be implemented within this approach using $\mathbb{Z}_2$-gradings or $\mathbb{Z}_2\times \mathbb{Z}_2$-gradings.  We will also see that a notion of {\em partial-grading} on the Lie algebra also emerges naturally in this context. 

There are two reasons why one might be interested in such purely algebraic constructions. One is practical. After the work of \cite{Krishnan:2013wta}, it was noticed that there exist a few contexts  were the calculations were easily implemented via the Grassman approach. The work of \cite{Krishnan:2013wta} lead to singularity resolutions of flat space cosmological solutions \cite{Krishnan:2013tza,Craps:2014wpa,Kiran:2014kca}. Another application was the addition of chemical potential in flat space higher spin theories where the Grassmann approach was used instrumentally in \cite{Gary:2014ppa}. In particular, the existence of a nilpotent parameter $\epsilon$ turned out to be handy for various algebraic and differential reasons in these papers. One might legitimately hope that the constructions we undertake here might lead to similar uses as well. 

Another reason for interest in this type of constructions is more fundamental. The usual analytic construction of contractions makes one somewhat more reliant on the underlying spacetime structures (like coordinates, derivatives, etc) than perhaps necessary.  It is curious then, that there exist purely algebraic relations relating physically interesting contractions. From a physics perspective, the existence of such an algebraic connection is perhaps a suggestion that the parallels that people often seek between {\em theories} related by such contractions should be understandable purely algebraically\footnote{Note that we are not suggesting that all aspects of the theories related by contractions are in 1-to-1 relation to each other. Our suggestion is that whatever aspects {\em are} related, are plausibly a consequence of algebraic facts alone.}.

In the rest of the paper, we will first introduce the general definitions of gradings first, and then show numerous examples that show that they subsume the various examples of contractions that have shown up in the literature. In an appendix, some closely related past work that might be useful for generalizing our work, is reviewed. The notion of partial grading on an algebra, which is a natural offshoot of our ideas is introduced in another appendix. To illustrate that not all contractions have to be based on $\mathbb{Z}_2$ or variations thereof, we discuss a slightly more exotic $\mathbb{Z}_7$ grading on an $su(3)$ algebra and its associated contraction in yet another appendix. It will be interesting to see whether our approach has generalizations or adaptations for super-algebras and also for algebras with non-linear structures like ${\cal W}_N$ algebras. It will also be interesting to see whether a calculus can be defined on the nilpotent structures we encounter, that generalize Berezin-like structures in the Grassmann case: this will have practical applications.


\section{Graded Lie Algebras: Generalities}\label{grading_gen}

We will start with a very brief review of gradings of algebras. One thing to keep in mind in what follows is that gradings on algebras are a basis dependent statement. 

Let $L$ be a Lie algebra (finite or infinite dimensional) and $(G,+)$ be an abelian group. The Lie algebra $L$ is said to be graded over $G$ if there exists a vector space decomposition,
\begin{eqnarray}
L = \bigoplus_{g \in G} L_g\;\;,
\end{eqnarray}
such that
\begin{eqnarray}
\left[T_a,T_b \right] \subseteq L_{g+h}, \quad \forall \quad T_a \in L_g, T_b \in L_h.
\end{eqnarray}
The elements $L_g$ are said to be \textit{homogeneous of degree} $g$. It can be immediately seen that $L_e$, where $e$ is the identity element of $G$, forms a subalgebra.

Graded algebras are familiar in the context of supersymmetry, but let us emphasize that supersymmetry is not necessary. A simple example of a graded Lie algebra is $\mathfrak{su}(2)$, which in the fundamental representation is generated by Pauli matrices

\begin{eqnarray}\label{pauli_matrices}
\tau_1 = \frac{1}{2} \left( \begin{array}{cc}
0 & 1 \\
1 & 0
\end{array} \right), \quad
\tau_2 = \frac{1}{2} \left( \begin{array}{cc}
0 & -i \\
i & 0
\end{array} \right), \quad
\tau_3 = \frac{1}{2} \left( \begin{array}{cc}
1 & 0 \\
0 & -1
\end{array} \right) 
\end{eqnarray}
There is a natural $\mathbb{Z}_2=\{0,1\}$ grading on the algebra given by

\begin{eqnarray}
L_0 = \{\tau_3\}, \quad L_1 = \{\tau_1, \tau_2\}
\end{eqnarray}

\section{Grading and $\epsilon$-Contractions}

We will start with gradings by $\mathbb{Z}_m$ group. There is a special kind of contraction possible when the grading group is $\mathbb{Z}_m$, which we will call $\epsilon$ \textit{contraction}, which can be implemented algebraically at the level of Lie algebra. For the $\mathbb{Z}_m$ graded Lie algebras, we note that the algebra has a decomposition
\begin{eqnarray}
L = \bigoplus_{\alpha = 0}^{m-1} L_{\alpha},
\end{eqnarray}
so that
\begin{eqnarray}
[T^{(\alpha)}_a,T^{(\beta)}_b] = f_{ab}^{\;\;\;c}\;\; T^{(\delta)}_c, \quad \delta \equiv (\alpha+\beta)\; \text{mod}\;m
\end{eqnarray}
The element $T^{\alpha}_{a}$ is said to be \textit{homogeneous of degree} $\alpha$. 

We will say that a $\mathbb{Z}_m$-graded Lie algebra is $\epsilon$ contracted if we replace every homogeneous element $T^{\alpha}_i$ by $T^{\alpha}_i \rightarrow \epsilon^{\alpha}T^{\alpha}_i$, where $\epsilon^{\alpha}$ are homogeneous elements of the quotient ring\footnote{A general element of this quotient ring is denoted \[a_0 + a_1 \epsilon + a_2 \epsilon^2 \cdots a_{m-1}\epsilon^{m-1} \] with the identification $\epsilon^m = 0$} $\mathbb{R}[\epsilon]/\epsilon^m$. This is a generalization of the Grassmann contraction of \cite{Krishnan:2013wta,Henkel:2005zu}. If we denote the contracted algebra by $\mathfrak{g}_c$, then
\begin{eqnarray}
\mathfrak{g}_c \subset \mathfrak{g} \otimes \mathbb{R}[\epsilon]/\epsilon^m
\end{eqnarray}
From the above definition we can see that all commutators $[T^{(\alpha)}_i,T^{(\beta)}_j]$ with $\alpha+\beta \geq m$ are contracted to zero in $\mathfrak{g}_c$.

It sould be noted that $\epsilon$-contraction leads to a Lie algebra. To establish this, all we need to show is that the contracted algebra satisfies Jacobi identity. If we denote the new generators to be 
\begin{eqnarray}
\mathcal{T}^{\alpha}_a := \epsilon^{\alpha}T^{\alpha}_a,
\end{eqnarray}
it then follows from the definition of $\epsilon$-contraction that if the original algebra satisfies Jacobi indentity, the contracted algebra will also satisfy the Jacobi identity:
\begin{eqnarray}\label{jacobiX}
[\mathcal{T}^{\alpha}_a,[\mathcal{T}^{\beta}_b,\mathcal{T}^{\gamma}_c]] + [\mathcal{T}^{\beta}_b,[\mathcal{T}^{\gamma}_c,\mathcal{T}^{\alpha}_a]] + [\mathcal{T}^{\gamma}_c,[\mathcal{T}^{\alpha}_a,\mathcal{T}^{\beta}_b]] &=& \epsilon^{\alpha+\beta+\gamma}\left( [T^{\alpha}_a,[T^{\beta}_b,T^{\gamma}_c]] \right. \\ \nonumber &+& \left. [T^{\beta}_b,[T^{\gamma}_c,T^{\alpha}_a]] + [T^{\gamma}_c,[T^{\alpha}_a,T^{\beta}_b]] \right)=0
\end{eqnarray}
We also observe an interesting feature of the $\epsilon$-contraction; the resulting algebra inherits the grading from the parent algebra. 

The above results can be generalized for grading by $\mathbb{Z}_{m_1}\times\mathbb{Z}_{m_2}\times \cdots \times \mathbb{Z}_{m_n}$. A general element of the group can be represented by an $n$-tuple $(\alpha_1,\alpha_2,\cdots,\alpha_n)$, where $\alpha_i \in (0,\dots,m_i-1)$ and the algebra can be contracted by introducing $n$ elements $(\epsilon_1,\epsilon_2,\cdots,\epsilon_n)$ such that
\begin{eqnarray}
\epsilon_{1}^{m_1} = \epsilon_{2}^{m_2} = \cdots \epsilon_{n}^{m_n} = 0.
\end{eqnarray}
The homogeneous generators of degree $(\alpha_1,\alpha_2,\cdots,\alpha_n)$ is thus redefined to be
\begin{eqnarray}
\mathcal{T}^{(\alpha_1,\alpha_2,\cdots,\alpha_n)}_a := \epsilon_{1}^{\alpha_1}\epsilon_{2}^{\alpha_2}\cdots \epsilon_{n}^{\alpha_n}T^{(\alpha_1,\alpha_2,\cdots,\alpha_n)}_a
\end{eqnarray}
which implements the $\epsilon$ contraction. 

A corollary of the fundamental theorem of Abelian groups is that every Abelian group is isomorphic to a group of the above form, therefore as far as gradings with Abelian groups go, we are done.

\section{Examples}\label{examples}

We will now present many examples of the general structure we have described, by implementing various contractions in the literature using our approach. We will see numerous examples of $\mathbb{Z}_2$-graded $\epsilon$-Contractions, one non-trivial example of a $\mathbb{Z}_2 \times \mathbb{Z}_2$ grading and one example of a $\mathbb{Z}_3$ grading. We will also see that different gradings on the parent algebra can lead to different resulting contracted algebras.

\subsection{(anti)-de Sitter to Poincare algebra}
The isometry group of a general d-dimensional de Sitter space is $SO(d,1)$ while the isometry group of anti-de Sitter group is $SO(d-1,2)$. The isometry algebra in both cases can be written as:

\begin{eqnarray}
[M^{AB},M^{CD}] = i(\eta^{AC}M^{BD} + \eta^{BD}M^{AC} - \eta^{BC}M^{AD} - \eta^{AD}M^{BC})
\end{eqnarray}
where $\eta = \text{diag}(-1,1,\cdots,1,1)$ for the de Sitter algebra and $\eta = \text{diag}(-1,1,\cdots,1,-1)$ for the anti-de Sitter algebra. The indices $A,B,\cdots$ runs from $0$ to $d$. To do the contraction, we seperate out the $x^d$ component of the algebra and the algebra can be now expressed as:

\begin{eqnarray}
[M^{d\mu},M^{d\nu}] &=& \pm i M^{\mu \nu} \\ \nonumber
[M^{d\mu},M^{\nu \rho}] &=& -i(\eta^{\mu \nu} M^{d\rho} - \eta^{\mu \rho}M^{d\nu}) \\ \nonumber
[M^{\mu \nu},M^{\rho \sigma}] &=& i(\eta^{\mu \rho}M^{\nu \sigma} + \eta^{\nu \sigma}M^{\mu \rho} - \eta^{\nu \rho}M^{\mu \sigma} - \eta^{\mu \sigma}M^{\nu \rho})
\end{eqnarray}
where $+$ and $-$ sign in the first equation refers to the de Sitter and anti-de Sitter algebra respectively. A little inspection shows that there is a $\mathbb{Z}_2$ grading on the algebra where the generators $M^{\mu \nu}$ are of degree $0$ and generators $M^{d \mu}$ are of degree $1$. To do an $\epsilon$ contraction of the above algebra, we define the new generator $P^{\mu}$ by
\begin{eqnarray}
P^{\mu} = \epsilon M^{d \mu}
\end{eqnarray}
and compute the commutators subject to $\epsilon^2 = 0$, which results in the Poincare algebra
\begin{eqnarray}
[P^{\mu},P^{\nu}]_{\e} &=& 0 \\ \nonumber
[P^{\mu},M^{\nu \rho}]_{\e} &=& -i(\eta^{\mu \nu} P^{\rho} - \eta^{\mu \rho}P^{\nu}) \\ \nonumber
[M^{\mu \nu},M^{\rho \sigma}]_{\e} &=& i(\eta^{\mu \rho}M^{\nu \sigma} + \eta^{\nu \sigma}M^{\mu \rho} - \eta^{\nu \rho}M^{\mu \sigma} - \eta^{\mu \sigma}M^{\nu \rho})
\end{eqnarray}

\subsection{$\mathfrak{su}(2)$ to $\mathfrak{iso}(2)$}
As an illustration of $\mathbb{Z}_3$ graded contraction, we look at the contraction of $\mathfrak{su}(2)$. $\mathfrak{su}(2)$ Lie algebra admits a $\mathbb{Z}_3$ grading, which is different from the one discussed in Sec. 2. To identify the grading, we go to the basis of lowering and raising operators where we have

\begin{eqnarray}
\tau^+ &=& \tau_1 + i \tau_2 \\ \nonumber
\tau^- &=& \tau_1 - i \tau_2 \\ \nonumber
\tau^0 &=& \tau_3 \\ \nonumber
\end{eqnarray}
where $\tau$s are the Pauli matrices, defined in \eqref{pauli_matrices}. The commutation relations now takes the form

\begin{eqnarray}
[\tau^0,\tau^{\pm}] = \pm \tau^{\pm},\qquad \quad [\tau^{+},\tau^{-}] = \tau^0
\end{eqnarray}
From the above commutation relation, it can be seen that there is an underlying $\mathbb{Z}_3$ group which can be identified as $\tau^0 \rightarrow 0$, $\tau^+ \rightarrow 1$ and $\tau^- \rightarrow 2$. In order to do the $\epsilon$ contraction, we define the new generators as

\begin{eqnarray}
J^0 = \tau^0,\quad P^+ = \epsilon \tau^+, \quad P^- = \epsilon^2 \tau^-
\end{eqnarray}
with $\epsilon^3=0$ and we obtain the new algebra

\begin{eqnarray}
[J^0, P^{\pm}] = \pm P^{\pm}, \qquad \quad [P^+, P^-] = 0
\end{eqnarray}
which is nothing but the $\mathfrak{iso}(2)$ algebra.

We would like to point it out that the $\mathbb{Z}_2$ grading on $\mathfrak{su}(2)$ which we identified in Sec. 2 can also be contracted. The result is, however, same as that of $\mathbb{Z}_3$ graded contraction.

\subsection{Galilean algebra}

Next we illustrate the example of an Inonu-Wigner contraction of Poincare algebra to Galilean algebra, first considered in \cite{Inonu:1953sp}. 

In terms of space $+$ time form, the Poincare algebra in $d$ dimensions can be written as:
\begin{eqnarray}
i[P^0,P^i] &=& 0,\qquad \qquad \qquad \qquad i[P^i,P^j] = 0, \\ \nonumber
i[P^0,M^{ij}] &=& 0, \qquad \qquad \qquad \qquad i[P^0,M^{0i}] = -P^i, \\ \nonumber
i[P^i,M^{jk}] &=& \delta^{ij}P^k - \delta^{ik}P^j, \quad \quad \;\;  i[P^i,M^{0j}] = -\delta^{ij}P^0, \\ \nonumber
i[M^{ij},M^{kl}] &=& \delta^{il}M^{jk} + \delta^{lk}M^{il} - \delta^{ik}M^{jl} - \delta^{jl}M^{ik} \\ \nonumber
i[M^{0i},M^{jk}] &=& \delta^{ij}M^{0k}  - \delta^{ik}M^{0j} \\ \nonumber
i[M^{0i},M^{0j}] &=& -M^{ij} \nonumber
\end{eqnarray}
We identify that there is a $\mathbb{Z}_2$ grading group where the generators $P^0$, $M^{ij}$ are of degree $0$ and $P^i$, $M^{0i}$ are of degree $1$. This is the grading which is relevant for contracting the Poincare to Galilean algebra. Explicitely, we define the new generators
\begin{eqnarray}
C^i \equiv \epsilon M^{0i}, \quad \qquad \mathcal{P}^i \equiv \epsilon P^i
\end{eqnarray} 
where $\epsilon^2=0$ is Grassmannian. The original Poincare algebra then reduces to
\begin{eqnarray}
i[E,\mathcal{P}^i] &=& 0,\qquad \qquad \qquad \qquad i[\mathcal{P}^i,\mathcal{P}^j] = 0, \\ \nonumber
i[E,L^{ij}] &=& 0, \qquad \qquad \qquad \qquad i[E,C^{i}] = \mathcal{P}^{i}, \\ \nonumber
i[\mathcal{P}^i,L^{jk}] &=& \delta^{ij}\mathcal{P}^k - \delta^{ik}\mathcal{P}^j, \quad \quad \;\;  i[\mathcal{P}^i,C^{j}] = 0 \\ \nonumber
i[C^{i},L^{jk}] &=& \delta^{ij}C^{k}  - \delta^{ik}C^{j}, \quad \quad \;\; i[C^{i},C^{j}] = 0 \\ \nonumber
i[L^{ij},L^{kl}] &=& \delta^{il}L^{jk} + \delta^{lk}L^{il} - \delta^{ik}L^{jl} - \delta^{jl}L^{ik} \nonumber
\end{eqnarray}
which is the Galilean algebra $G(d-1,1)$ without the central extension.

Poincare algebra is an example of the (unsurprising) fact that grading structures on algebras need not be unique, even with the same grading group. One can see that there is another $\mathbb{Z}_2$ grading group, where $P^i$ and $M^{ij}$ are of degree $0$ whereas $P^0$ and $M^{0i}$ are of degree $1$, acting on the Pincare algebra. However, the graded contraction of this $\mathbb{Z}_2$ does not give the Galilean algebra. Following our recipe, we identify new generators
\begin{eqnarray}
E = \epsilon P^0,\quad C^i = \epsilon M^{0i}, \quad L^{ij} = M^{ij},
\end{eqnarray}
with the condition $\epsilon^2 =0$. The new algebra is given by,
\begin{eqnarray}
i[E,P^i] &=& 0,\qquad \qquad \qquad i[P^i,P^j] = 0, \\ \nonumber
i[E,M^{ij}] &=& 0, \qquad \qquad \qquad i[E,C^{i}] = 0, \\ \nonumber
i[P^i,L^{jk}] &=& \delta^{ij}P^k - \delta^{ik}P^j, \quad i[P^i,C^{j}] = -\delta^{ij}E \\ \nonumber
i[L^{ij},L^{kl}] &=& \delta^{il}L^{jk} + \delta^{lk}L^{il} - \delta^{ik}L^{jl} - \delta^{jl}L^{ik} \\ \nonumber
i[C^{i},M^{jk}] &=& \delta^{ij}C^{k}  - \delta^{ik}C^{j} \\ \nonumber
i[C^{i},C^{j}] &=& 0 \nonumber
\end{eqnarray}
Even though this algebra does not seem to have occured in the literature, we feel that it might be of some interest because the grading we start with seems like a natural one to consider. We will not discuss it further. 

\subsection{Galilean Conformal Algebra}
It was shown in \cite{Bagchi:2009my,Martelli:2009uc} that the non-relativistic limit of the relativistic conformal algebra in $d$ dimension is given by the Galilean conformal algebra (GCA). This can be thought of as an extension of the contraction from Poincare to Galilean algebra as discussed in the previous section. In what follows, we will adopt the conventions of \cite{Bagchi:2009my}, and the relativistic conformal algebra in $d$ dimensions can be written as
\begin{eqnarray}
[L_{ij},H] &=& 0 \quad \quad \quad [P_i,P_j] = 0, \quad \quad [C_i,C_j] = -L_{ij} \\ \nonumber
[C_i,P_j] &=& \delta_{ij}H, \quad [H,P_i] = 0,\quad \quad [H,C_i] = -P_i \\ \nonumber
[L_{ij},P_k] &=& -\left(\delta_{jk}P_i - \delta_{ik}P_j \right), \qquad  [L_{ij},C_k] = -\left(\delta_{jk}C_i - \delta_{ik}C_j \right) \\ \nonumber
[L_{ij},L_{kl}] &=& -(\delta_{il}L_{jk} + \delta_{lk}L_{il} - \delta_{ik}L_{jl} - \delta_{jl}L_{ik}) \\ \nonumber
[K,K_i] &=& 0, \quad \quad \;\;\; [K,C_i] = K_i, \quad \quad [K,P_i] = 2C_i \\ \nonumber
[L_{ij},K_k] &=& -\left(\delta_{jk}K_i - \delta_{ik}K_j \right), \quad [L_{ij},K] = 0, \quad [L_{ij},D] = 0 \\ \nonumber
[K_i,K_j] &=& 0, \quad \quad [K_i,C_j] = \delta_{ij}K, \quad [K_i,P_j] = 2L_{ij}+2\delta_{ij}D \\ \nonumber
[H,K_i] &=& -2C_i \quad \quad [D,K_i] = -K_i, \quad \quad [D,C_i] = 0, \\ \nonumber
[D,P_i] &=& P_i, \quad [D,H] = H, \quad [H,K] = -2D, \\ \nonumber
[D,K] &=& -K
\end{eqnarray}
where we have made explicite space $+$ time split. Apart from the Poincare generators, it contains the conformal generators- $D$ is the dilatation generator, $K$ is the $0$-component of the special conformal generators and $K_i$ are its spatial components. The requisite grading group which is relevant for contraction to GCA is $\mathbb{Z}_2$ and we identify them as: $\{L_{ij}, H, D, K\}$ are of degree $0$ while $\{C_i,P_i,K_i\}$ are of degree $1$. Having identified the grading group, now we may introduce the new generators 
\begin{eqnarray}
\mathcal{P}_i \equiv \epsilon P_i,\quad \mathcal{C}_i \equiv \epsilon C_i, \quad \mathcal{K}_i \equiv \epsilon K_i
\end{eqnarray}
where $\epsilon$ is Grassmannian. The commutator of the redefined generators give the required contracted algebra GCA
\begin{eqnarray}
[L_{ij},H] &=& 0 \quad \quad \quad [\mathcal{P}_i,\mathcal{P}_j] = 0, \quad \quad [\mathcal{C}_i,\mathcal{C}_j] = 0 \\ \nonumber
[\mathcal{C}_i,\mathcal{P}_j] &=& 0, \quad [H,\mathcal{P}_i] = 0,\quad \quad [H,\mathcal{C}_i] = -\mathcal{P}_i \\ \nonumber
[L_{ij},\mathcal{P}_k] &=& -\left(\delta_{jk}\mathcal{P}_i - \delta_{ik}\mathcal{P}_j \right), \qquad  [L_{ij},\mathcal{C}_k] = -\left(\delta_{jk}\mathcal{C}_i - \delta_{ik}\mathcal{C}_j \right) \\ \nonumber
[L_{ij},L_{kl}] &=& -\left( \delta_{il}L_{jk} + \delta_{lk}L_{il} - \delta_{ik}L_{jl} - \delta_{jl}L_{ik} \right) \\ \nonumber
[K,\mathcal{K}_i] &=& 0, \quad \quad \;\;\; [K,\mathcal{C}_i] = \mathcal{K}_i, \quad \quad [K,\mathcal{P}_i] = 2\mathcal{C}_i \\ \nonumber
[L_{ij},\mathcal{K}_k] &=& -\left(\delta_{jk}\mathcal{K}_i - \delta_{ik}\mathcal{K}_j \right), \quad [L_{ij},K] = 0, \quad [L_{ij},D] = 0 \\ \nonumber
[\mathcal{K}_i,\mathcal{K}_j] &=& 0, \quad \quad [\mathcal{K}_i,\mathcal{C}_j] = 0, \quad [\mathcal{K}_i,\mathcal{P}_j] = 0 \\ \nonumber
[H,\mathcal{K}_i] &=& -2\mathcal{C}_i \quad \quad [D,\mathcal{K}_i] = -\mathcal{K}_i, \quad \quad [D,\mathcal{C}_i] = 0, \\ \nonumber
[D,\mathcal{P}_i] &=& \mathcal{P}_i, \quad [D,H] = H, \quad [H,K] = -2D, \\ \nonumber
[D,K] &=& -K
\end{eqnarray}

\subsection{$\mathfrak{Vir}\oplus \mathfrak{Vir}$ to $\mathfrak{bms}_3$}

In AdS$_3$ with Brown-Henneaux boundary condition, the asymptotic symmetry algebra is given by two copies of Virasoro algebra with non-zero central charge, $c=3l/2G_N$. In \cite{Krishnan:2013wta}, it was shown that implementing the Inonu-Wigner contraction using a Grassmann parameter maps the space of solutions, asymptotic charges and action to those of asymptotically flat space. Here we show the equivalence of that approach to a graded contraction of the Virasoro algebra. 

We begin with the two copies of Virasoro algebra 
\begin{eqnarray}
[L^{\pm}_m,L^{\pm}_n] = (m-n)L^{\pm}_{m+n} + \frac{c^{\pm}}{12}m (m^2-1)\delta_{m+n,0}, \quad [L^{+}_m,L^{-}_m] = 0.
\end{eqnarray}
We can combine the two set of generators into 2 set of linear combinations that has a natural $\mathbb{Z}_2=\{0,1\}$ grading by
\begin{eqnarray}
L^{0}_n = L^{+}_n - L^{-}_{-n},\quad L^{1}_n = L^{+}_n + L^{-}_{-n}
\end{eqnarray}
The Virasoro algebra in this basis takes the form

\begin{eqnarray}
[L^{0}_m,L^{0}_n] &=& (m-n)L^{0}_{m+n},\\ \nonumber
[L^{1}_m,L^{1}_n] &=& (m-n)L^{0}_{m+n},\\ \nonumber
[L^{0}_m,L^{1}_n] &=& (m-n)L^{1}_{m+n} + \frac{C}{12}m(m^2 -1)\delta_{m+n,0} \nonumber
\end{eqnarray}
where $C = c^+ + c^-$ and the central element is identified to be of degree 1. The $\epsilon$ contraction of the above algebra is achieved by redefining the generators of degree 1 by multiplying them with a Grassmann parameter $\epsilon$ such that $L^{1}_m \rightarrow \tilde L^{1}_m= \epsilon L^{1}_n$ and $C \rightarrow \tilde C = \epsilon C$, and was first shown in \cite{Krishnan:2013wta}.
\begin{eqnarray}
[L^{0}_m,L^{0}_n] &=& (m-n)L^{0}_{m+n},\\ \nonumber
[L^{1}_m,L^{1}_n] &=& 0,\\ \nonumber
[L^{0}_m,L^{1}_n] &=& (m-n)L^{1}_{m+n} + \frac{C}{12}m(m^2 -1)\delta_{m+n,0}
\end{eqnarray}
In writing the final result, we have suppressed the tilde's. 
The resulting algebra is the flat space asymptotic symmetry algebra $\mathfrak{bms}_3$, \cite{Barnich:2006av}. In \cite{Krishnan:2013wta}, it was further shown that $\epsilon$ contractions can be used to contract the spin-3 algebra in AdS$_3$ to flat space. It is easy to check following our present discussion that there is a $\mathbb{Z}_2$ grading group which gives this higher spin contraction, but we will not present the details here explicitly.


\subsection{$\mathfrak{sl}(2)_k \oplus \mathfrak{sl}(2)_k$ to $\mathfrak{isl}(2)_k$}
The ASA of the loosest set of AdS$_3$ boundary condition is given by $\mathfrak{sl}(2)_k \oplus \mathfrak{sl}(2)_k$ \cite{Grumiller:2016pqb}. Under the Inonu-Wigner contraction this algebra is known to map to an $\mathfrak{isl}(2)_k$ which is the ASA of most general flat space solution. Here we show that the Inonu-Wigner contraction can be realized as a $\mathbb{Z}_2$ graded contraction of $\mathfrak{sl}(2)_k \oplus \mathfrak{sl}(2)_k$ algebra similar to the case of Virasoro algebra of previous section.

$\mathfrak{sl}(2)_k$ Kac-Moody algebra is given by generators $\mathfrak{J}^{a}_{n}$ satisfying the commutation relation

\begin{eqnarray}\label{sl2k_algebra}
[\mathfrak{J}^{a}_{m},\mathfrak{J}^{b}_{n}] = (a-b)\mathfrak{J}^{a+b}_{m+n} - \frac{1}{2}n \hat{k} \kappa_{ab} \delta_{m+n,0}.
\end{eqnarray}
where $a,b=-1,0,1$ and $m,n \in \mathbb{Z}$. As in the previous case, we form linear combinations which have $\mathbb{Z}_2$ grading

\begin{eqnarray}
J^{a}_n = \mathfrak{J}^{a}_{n} - \bar{\mathfrak{J}}^{-a}_{-n}, \quad P^{a}_n = \mathfrak{J}^{a}_{n} + \bar{\mathfrak{J}}^{-a}_{-n}
\end{eqnarray} 
$\mathfrak{sl}(2)_k$ algebra now takes the form

\begin{eqnarray}
[J^{a}_n, J^{b}_m] &=& (a-b)J^{a+b}_{n+m}, \\ \nonumber
[P^{a}_n, P^{b}_m] &=& (a-b)J^{a+b}_{n+m}, \\ \nonumber
[J^{a}_n, P^{b}_m] &=& (a-b)P^{a+b}_{n+m} - n \hat{k} \kappa_{ab} \delta_{n+m,0}. \nonumber
\end{eqnarray}
where it can be seen that the generators $J^{a}_n$ are of degree $0$ and $P^{a}_n$ are of degree $1$. The central element $\hat{k}$ once again is of degree $1$.To obtain the $\mathfrak{isl}(2)_k$ algebra we make the replacement $P^{a}_n \rightarrow \epsilon P^{a}_n$ and $\hat{k} \rightarrow k \equiv \epsilon \hat{k}$, where $\epsilon^2 =0$. The resulting algebra is

\begin{eqnarray}
[J^{a}_n, J^{b}_m] &=& (a-b)J^{a+b}_{n+m}, \\ \nonumber
[P^{a}_n, P^{b}_m] &=& 0, \\ \nonumber
[J^{a}_n, P^{b}_m] &=& (a-b)P^{a+b}_{n+m} - n k \kappa_{ab} \delta_{n+m,0}.
\end{eqnarray}

\subsection{Carroll Algebra from $\mathfrak{sl}(2)_k \oplus \mathfrak{sl}(2)_k$}
In \cite{Grumiller:2017sjh} it was shown that $\mathfrak{sl}(2)_k \oplus \mathfrak{sl}(2)_k$ has another interesting Inonu-Wigner contraction which maps it the loop algebra of Carroll gravity. This serves as a good example that can be understood as a $\mathbb{Z}_2 \times \mathbb{Z}_2$ graded  $\epsilon$ contraction. 

$\mathfrak{sl}(2)_k$ algebra given by \eqref{sl2k_algebra} is isomorphic to $\mathfrak{so}(2,1)_k$ and this identification helps us to identify a semi-grading on the algebra. $\mathfrak{so}(2,1)_k$ is given by \cite{Grumiller:2017sjh}
\begin{eqnarray}
[\mathfrak{J}^{a}_{m},\mathfrak{J}^{b}_{n}] = \epsilon^{ab}_{\;\;\;c} \mathfrak{J}^{c}_{m+n} - \frac{1}{2}n \hat{k} \eta^{ab} \delta_{m+n,0}.
\end{eqnarray}  
where $\eta_{ab} = \text{diag}(-1,1,1)$ is the $(2+1)$ -dimensional Minkowski metric and the three dimensional Levi-Civit\'a symbol is taken with the convention $\epsilon_{012} = 1$, $\epsilon_{ij} = \epsilon_{0ij}$. We now distinguish the $a=0$ component from the spatial components $a=i$, $i=1,2$ and form new generators
\begin{eqnarray}
J_m &=& -(\mathfrak{J}^{0}_{m} + \bar{\mathfrak{J}}^{0}_{-m}), \\ \nonumber 
P^{i}_m &=& \mathfrak{J}^{i}_{m} + \bar{\mathfrak{J}}^{i}_{-m}, \\ \nonumber
G^{i}_m &=& \mathfrak{J}^{i}_{m} - \bar{\mathfrak{J}}^{i}_{-m}, \\ \nonumber
H_m &=& -(\mathfrak{J}^{0}_{m} - \bar{\mathfrak{J}}^{0}_{-m})
\end{eqnarray}
The non-zero commutators can be written as:
\begin{eqnarray}\label{sl2ktocarroll}
[J_m,P^{i}_n] &=& \epsilon^{i}_{\;\;j}P^{j}_{m+n} \\ \nonumber
[J_m,G^{i}_n] &=& \epsilon^{i}_{\;\;j}G^{j}_{m+n} \\ \nonumber
[J_m,H_n] &=& \hat{k} m \delta_{m+n,0} \\ \nonumber
[P^{i}_m,G^{j}_n] &=& -\epsilon^{ij}H_{m+n} - \hat{k} m \delta^{ij} \delta_{m+n,0}\\ \nonumber
[P^{i}_m,H_n] &=& -\epsilon^{i}_{\;\;j}G^{j}_{m+n} \\ \nonumber
[G^{i}_m,H_n] &=& -\epsilon^{i}_{\;\;j}P^{j}_{m+n}  \nonumber
\end{eqnarray}
From the commutation relations above, it can be seen that there is a $\mathbb{Z}_2 \times \mathbb{Z}_2$ grading group which provides an exact grading of the algebra. The identification of the generators with the grading group is as follows:
\begin{eqnarray}
J_n & \rightarrow & (0,0) \\ \nonumber
P^i_n & \rightarrow & (1,0) \\ \nonumber
G^i_n & \rightarrow & (0,1) \\ \nonumber
H_n & \rightarrow & (1,1)  \nonumber
\end{eqnarray}
To contract the above algebra to the Carroll loop algebra, we can introduce two Grassmann parameters $\xi$ and $\eta$ and redefine the generators
\begin{eqnarray}
\mathcal{P}^i_n \equiv \xi P^i_n,\quad \mathcal{G}^i_n \equiv \eta G^i_n, \quad \mathcal{H}_n \equiv \xi \eta H_n, \quad k \equiv \xi \eta \hat{k}
\end{eqnarray}
such that $\xi^2 = 0$, $\eta^2 =0$. With these identifications, the commutation relation \eqref{sl2ktocarroll} takes the form
\begin{eqnarray}
[J_m,\mathcal{P}^{i}_n] &=& \epsilon^{i}_{\;\;j}\mathcal{P}^{j}_{m+n} \\ \nonumber
[J_m,\mathcal{G}^{i}_n] &=& \epsilon^{i}_{\;\;j}\mathcal{G}^{j}_{m+n} \\ \nonumber
[J_m,\mathcal{H}_n] &=& k m \delta_{m+n,0} \\ \nonumber
[\mathcal{P}^{i}_m,\mathcal{G}^{j}_n] &=& -\epsilon^{ij}\mathcal{H}_{m+n} - k m \delta^{ij} \delta_{m+n,0} \nonumber
\end{eqnarray}
where all the other commutators vanish.

\section*{Acknowledgments}

We thank Arjun Bagchi for a discussion and Daniel Grumiller for a correspondence. 

\appendix

\section{Graded Contractions of Lie Algebras}

Inonu-Wigner contractions of the Lie algebra amounts to deforming the structure constants of the algebra, usually realized as transformations controlled by a parameter, which becomes singular at some point in the parameter space. In this appendix, we present a construction that is closely related to our work, due to \cite{deMontigny:1994nq, deMontigny:1995xb,Francisco-1,Francisco-2,Francisco-3}\footnote{This interesting work seems to be essentially unknown according to INSPIRE, and we stumbled upon it after we devised our own $\epsilon$-contractions approach. We have included a short review here in the hope that perhaps this could be used as a way to further work. Note that our approach contains a nilpotent parameter, and it played a crucial role in the physical applications considered in \cite{Krishnan:2013wta} as well as in \cite{Gary:2014ppa}.} which defines \textit{graded contractions} of the Lie algebra $L$ by introducing parameters $\e(g,h)$ such that the contracted algebra is given by the modified commutation relations
\begin{eqnarray}
[L_g,L_h]_{\e} \equiv \e(g,h)[L_g,L_h]
\end{eqnarray} 
Note that here, $g, h$ are elements of a general Abelian group.
It then follows from the Jacobi identity, that these parameters have to satisfy
\begin{eqnarray}\label{jacobi_contraint}
\e(g,h)\e(g+h,i) = \e(h,i)\e(i+h,g) = \e(g,i)\e(g+i,h)
\end{eqnarray}
In terms of $\e(g,h)$, for $\mathbb{Z}_m$ grading group, our $\epsilon$ grading is related to the choice 
\begin{eqnarray}
\e(\alpha,\beta) = \begin{cases} 
1 & \alpha+\beta < m \\
0 & \alpha+\beta \geq m \\ 
\end{cases}
\end{eqnarray}
As an explicit example, the $\mathbb{Z}_2$ $\epsilon$-contraction is related to the choice of $\e(\alpha,\beta)$ where
\begin{eqnarray}
\e(0,0) &=& 1, \\ \nonumber
\e(1,1) &=& 0, \\ \nonumber
\e(0,1) &=& 1 = \e(1,0) \nonumber
\end{eqnarray} 
It can be checked that the above choice is consistent with \eqref{jacobi_contraint}.

\section{$\mathbb{Z}_m$-Partial Grading and $\epsilon$-Contractions}\label{semi-grading}

Let $\mathfrak{g}$ be a Lie algebra with generators $\{T_1,T_2,\cdots,T_n\}$. We will call the algebra $\mathfrak{g}$ \textit{partially graded} of order $m$ if there exist a vector space decomposition
\begin{eqnarray}
\mathfrak{g} = \bigoplus_{\alpha = 0}^{m-1}\mathfrak{g}_{\alpha}, 
\end{eqnarray}
such that
\begin{eqnarray}
[T^{(\alpha)}_i,T^{(\beta)}_j] = f_{ij}^{\;\;\;k}T^{(\alpha+\beta)}_k, \quad \forall \quad (\alpha,\beta),\;\;\; \alpha + \beta < m
\end{eqnarray}
For $\alpha+\beta \geq m$, there is no grading and the commutator can close on any element of the Lie algebra. Therefore, partial-graded Lie algebras has roughly the structure of a $\mathbb{Z}_m$ grading as long as $\alpha+\beta < m$. It can be now seen that all $\mathbb{Z}_m$ graded Lie algebra are also partial-graded and $\mathfrak{g}_0$ is again a subalgebra. 

To implement $\epsilon$-contraction, we do not need a full grading, but only a partial grading, and this is why we are interested in it. Given a partially graded Lie algebra, the Inonu-Wigner contraction of the algebra can be implemented by the $\epsilon$-contraction. As the construction shows, partial-grading is the \emph{minimal} structure that ensures that the given Lie algebra can be $\epsilon$ contracted in this way\footnote{Note the interesting fact that in principle one can contract more general algebras than Lie algebras, and yet potentially end up with Lie algebras! This is because as longs as $\alpha+\beta+\gamma>m$, we have $\epsilon^{\alpha+\beta+\gamma}=0$ which guarantees that the Jacobi identity on the left hand side is automatically satisfied, even if the Jacobi-identity within the parenthesis on the RHS of \eqref{jacobiX} is not identically zero. So parent generators for which $\alpha+\beta+\gamma>m$ do not have to satisfy the Jacobi identity, for the corresponding descendant generators to satisfy it.}.

The $\epsilon$ contraction of the partial-graded algebras proceed along the same lines as that of $\mathbb{Z}_m$ graded Lie algebras. First, to each homogeneous element of the algebra, we assign homogeneous elements of $\mathbb{R}[\epsilon]/\epsilon^m$ and define the new generators
\begin{eqnarray}
\mathcal{T}^{\alpha}_a := \epsilon^{\alpha}T^{\alpha}_a,
\end{eqnarray}
The resulting Lie algebra of the redefined generators gives the $\epsilon$ contracted Lie algebra. It can also be checked that the contracted algebra is indeed a Lie algera by computing the Jacobi identity, which is trivially satisfied.

\section{$\mathbb{Z}_7$ grading on $\mathfrak{su}(3)$ }
So far all the physically interesting examples presented include gradings by $\mathbb{Z}_2$, $\mathbb{Z}_2 \times \mathbb{Z}_2$ and (a fairly trivial case of) $\mathbb{Z}_3$. Just to emphasize that we can have more non-trivial gradings, here we present the example of $\mathfrak{su}(3)$ Lie algebra which has a $\mathbb{Z}_7$ grading when written in the Cartan-Weyl basis. Similar constructions exist for other Lie algebras in the Cartan classification as well. Let us also note that if we contract them using our grading approach, the results seem to generate lots of new algebras. It will be interesting to see if they are of any practical utility.

In the Cartan-Weyl basis, the generators are denoted $\{t_{\pm},t_z,u_{\pm},v_{\pm},y\}$, where $t_z$ and $y$ form the Cartan subalgebra. The commutation relations are:
\begin{eqnarray}
[t_+,t_-] &=& 2t_z, \quad [u_+,u_-] = 2\left(\frac{3}{4}y - \frac{1}{2}t_z \right), \quad [v_+,v_-] = 2\left(\frac{3}{4}y + \frac{1}{2}t_z \right), \\ \nonumber
[t_+,u_+] &=& v_+, \quad [t_+,v_-]\;\; = -u_-, \quad [u_+,v_-]= t_-,\quad [v_+,t_-] = -u_+, \quad [v_+,u_-] = t_+, \\ \nonumber
[t_z,t_{\pm}] &=& \pm t_{\pm}, \quad [t_z,u_{\pm}] = \mp \frac{1}{2}u_{\pm}, \quad [t_z,v_{\pm}] = \pm \frac{1}{2}v_{\pm}, \\ \nonumber
[y,t_{\pm}] &=& 0, \quad \quad \;\; [y,u_{\pm}] = \pm u_{\pm}, \quad \;\;\; [y,v_{\pm}] = \pm v_{\pm}
\end{eqnarray}
From the above commutation relations we identify the $\mathbb{Z}_7$ grading to be $t_z,y \rightarrow 0$, $t_+ \rightarrow 1$, $u_+ \rightarrow 2$, $v_+ \rightarrow 3$, $v_- \rightarrow 4$, $u_- \rightarrow 5$ and $t_- \rightarrow 6$.

To implement the $\epsilon$ contraction, we define new generators
\begin{eqnarray}
T_+ = \epsilon t_+, \quad U_+ = \epsilon^2 u_+, \quad V_+ = \epsilon^3 v_+, \\ \nonumber
V_- = \epsilon^4 v_-, \quad U_- = \epsilon^5 u_-, \quad T_- = \epsilon^6 t_-,
\end{eqnarray}
and the commutation relations now become
\begin{eqnarray}
[T_+,T_-] &=& 0, \quad [U_+,U_-] = 0, \quad [V_+,V_-] = 0, \quad [T_+,U_+] = V_+, \\ \nonumber
[T_-,U_-] &=& 0, \quad [T_+,V_-] = -U_-, \quad [U_+,V_-] = T_-, \\ \nonumber
[V_+,T_-] &=& 0, \quad [V_+,U_-] = 0, \\ \nonumber
[T_z,T_{\pm}] &=& \pm T_{\pm}, \quad [T_z,U_{\pm}] = \mp \frac{1}{2}U_{\pm}, \quad [T_z,V_{\pm}] = \pm \frac{1}{2}V_{\pm}, \\ \nonumber
[Y,T_{\pm}] &=& 0, \qquad \;\;\; [Y,U_{\pm}] = \pm U_{\pm}, \quad \quad [Y,V_{\pm}] = \pm V_{\pm},
\end{eqnarray}
which seems to be a new algebra. Notice that $\mathfrak{su}(3)$ also admits a $\mathbb{Z}_2$ grading where the two Cartan generators are of grade $0$ and all the other generators are of grade $1$. We will not do the contraction for this grading here, but point out that this gives a different algebra.

\end{document}